\documentclass[10pt,twocolumn,superscriptaddress,english,10pt,prl,showpacs,floatfix,aps]{revtex4-1}
\usepackage{graphicx}
\usepackage{color}
\usepackage{epstopdf}
\usepackage{here}
\usepackage[utf8]{inputenc}
\usepackage{amsthm}
\usepackage{amsmath}
\usepackage{amssymb}
\usepackage{esint}
\usepackage{graphicx}
\usepackage{upgreek}
\usepackage{xspace}
\usepackage{ulem}
\usepackage{nicefrac}
\usepackage[separate-uncertainty=true]{siunitx}

\usepackage{soul}
\usepackage{braket}
\usepackage[backref=false,
bookmarksnumbered=true,
bookmarks=true,
bookmarksopen=true,
colorlinks=true,
citecolor=blue,
linkcolor=blue,
anchorcolor=green,
urlcolor=blue,unicode=false]{hyperref}

\renewcommand{\v}[1]{\ensuremath{\mathbf{#1}}} 
 
\let\baraccent=\= 
\renewcommand{\=}[1]{\stackrel{#1}{=}} 

\newcommand{\angstrom}{\textup{\AA}}

\newcommand{\mos}{MoS$_2$}

\newcommand{\iv}{d$I$/d$V$}
\newcommand{\zv}{$(\partial I$/$\partial V)_I$}
\newcommand{\iz}{d$I$/d$z$}
\newcommand{\kp}{$\bar K$}
\newcommand{\gp}{$\bar \Gamma$}
\newcommand{\qp}{$\bar Q$}
\newcommand{\mpoint}{$\bar M$}

\begin{document}

\title{Moir\'e structure of \mos\ on Au(111): Local structural and electronic properties}


\author{Nils Krane}
\affiliation{\mbox{Fachbereich Physik, Freie Universit\"at Berlin, Arnimallee 14, 14195 Berlin, Germany}}

\author{Christian Lotze}
\email{c.lotze@fu-berlin.de}
\affiliation{\mbox{Fachbereich Physik, Freie Universit\"at Berlin, Arnimallee 14, 14195 Berlin, Germany}}

\author{Katharina J. Franke}
\affiliation{\mbox{Fachbereich Physik, Freie Universit\"at Berlin, Arnimallee 14, 14195 Berlin, Germany}}

\date{\today}

\begin{abstract}
{Monolayer islands of molybdenum disulfide (MoS$_2$) on Au(111)} form a characteristic moir\'e structure, leading to locally different stacking sequences at the S--Mo--S--Au interface. Using low-temperature scanning tunneling microscopy (STM) and atomic force microscopy (AFM), we find that the moir\'e islands exhibit a unique orientation with respect to the Au crystal structure. This indicates a clear preference of MoS$_2$ growth in a regular stacking fashion. We further probe the influence of the local atomic structure on the electronic properties. Differential conductance spectra show pronounced features of the valence band and conduction band, some of which undergo significant shifts depending on the local atomic structure. We also determine the tunneling decay constant as a function of the bias voltage by a height-modulated spectroscopy method. This allows for an increased sensitivity of states with non-negligible parallel momentum $k_\parallel$ and the identification of the origin of the states from different areas in the Brillouin zone.
\end{abstract}

\maketitle 

\section{Introduction}

Transition metal dichalcogenides are a very interesting class of material with quasi-two-dimen\-sional properties due to their van der Waals-layered structure.
Molybdenum disulfide (\mos), which is a semiconductor when grown in the 2H-structure, belongs to this class of materials, with each sheet consisting of a Mo atomic layer sandwiched between two S planes as shown in figure \ref{fig:01}a.
Whereas the bulk crystal exhibits an indirect band gap, a single layer has a direct band gap located at the \kp\ point of the Brillouin zone (BZ)~\cite{Mak,Splendiani}.
The optical gap amounts to  $\approx$\SI{1.8}{\electronvolt}~\cite{Mak,Splendiani}, which covers the energy range of the visible spectrum and renders it interesting for optoelectronic devices.
The electronic band gap is larger by the exciton binding energy~\cite{Tawinan,Ashwin,Komsa,Qiu,Ugeda}.

Single-layer (SL) \mos\ can be obtained by exfoliation methods or by growing it directly on a variety of surfaces, providing large islands of high quality~\cite{Gronborg, Sorensen,Shi,Lee,Zhan,Wang}.
When the \mos\ layer is in contact with a metal surface, the band structure is strongly modified, due to screening~\cite{Roesner} and hybridization effects~\cite{Sorensen,Bruix,Dedzik}.
Bruix et al. have shown that the hybridization with a Au surface most strongly affects the states at the \gp\ point due to their out-of-plane character.
In contrast, the states at the \kp\ point remain largely unperturbed.  This can be explained by their in-plane character~\cite{Bruix} and the fact that the \kp\ point of \mos\ lies in a projected bulk band gap of the Au substrate~\cite{Miwa,Cabo}.

On Au(111) the monolayer \mos\ islands exhibit a lattice mismatch with the underlying surface and, thus, form a moir\'e superstructure~\cite{Bruix,Sorensen}.
One may expect that the local variations in atomic structure at the interface may lead to different strengths in the hybridization and screening effects.
The moir\'e structure would thus not only show topographic height variations but also periodic changes in the electronic structure.
Local measurement techniques, such as scanning tunneling microscopy and spectroscopy (STM/STS) are ideally suited to study the interplay of atomic-scale structure and electronic properties.
However, the drawback of STM is the lack of $k$-space resolution.
This challenge has recently been partially overcome by a measurement technique introduced by Zhang \textit{et al.}~\cite{Zhang}.
Detecting the bias-dependent lock-in signal from an oscillating tip allows to extract the tunneling decay constant, which entails information on the parallel momentum of the tunneling electron.

After characterizing the stacking of the \mos\ layer in detail, we use this technique to study the local electronic properties of \mos\ on Au(111). We show which states are affected by the different local interface structure with Au, and which states remain rather unaffected.

\section{Experimental details}

The Au(111) substrate was cleaned by repeated sputtering and annealing cycles under ultrahigh vacuum conditions.
For the growth of monolayer \mos\ islands, we adapted a recipe from Gr{\o}nborg et al~\cite{Gronborg}.
Molybdenum was deposited for $5-\SI{8}{\minute}$ in an H$_2$S atmosphere of \SI{e-5}{\milli\bar} onto the clean Au(111) surface by electron-beam evaporation from a high-purity rod.
To obtain \mos\ islands covering less than half of the Au(111) substrate, different evaporation rates of Mo were used, all below \SI{1} monlayer per hour.
Subsequently, the sample was annealed for 30\,min at \SI{550}{\celsius} in H$_2$S atmosphere.
Afterwards, the sample was transferred without exposing it to air into a CreaTec scanning tunneling microscope (STM) with a base temperature of \SI{4.6}{\kelvin}.
The microscope is equipped with a qPlus tuning fork sensor for combined measurements of tunneling and atomic force microscopy~\cite{Giessibl}.
Scanning tunneling spectroscopy (STS) was performed using a lock-in amplifier ($V_\mathrm{mod}=\SI{5.5}{\milli\volt}$, $f=\SI{921}{\hertz}$).
The STM tip was formed by controlled indentation into the surface and its quality assured by reference STS on the bare Au(111) surface.

In order to assess our results of the \mos\ electronic structure, we performed simple band structure calculations with the Quantum Espresso 6.0 package~\cite{QE} of freestanding single layer \mos.
Valence electrons were described by plane wave basis sets with \SI{408}{\electronvolt} kinetic energy cut off and the Projector Augmented Waves (PAW) method with the PBE approximation to the exchange-correlation functional were used.
Our results are consistent with other SL-\mos\ band structures from the literature~\cite{Bruix,Qiu,Padilha}.

\section[Single-layer MoS2 islands on Au(111)]{Single-layer \mos\ islands on Au(111)}
\label{islands}

\begin{figure*}[ht]
	\begin{center}
		\includegraphics[width=0.8\textwidth]{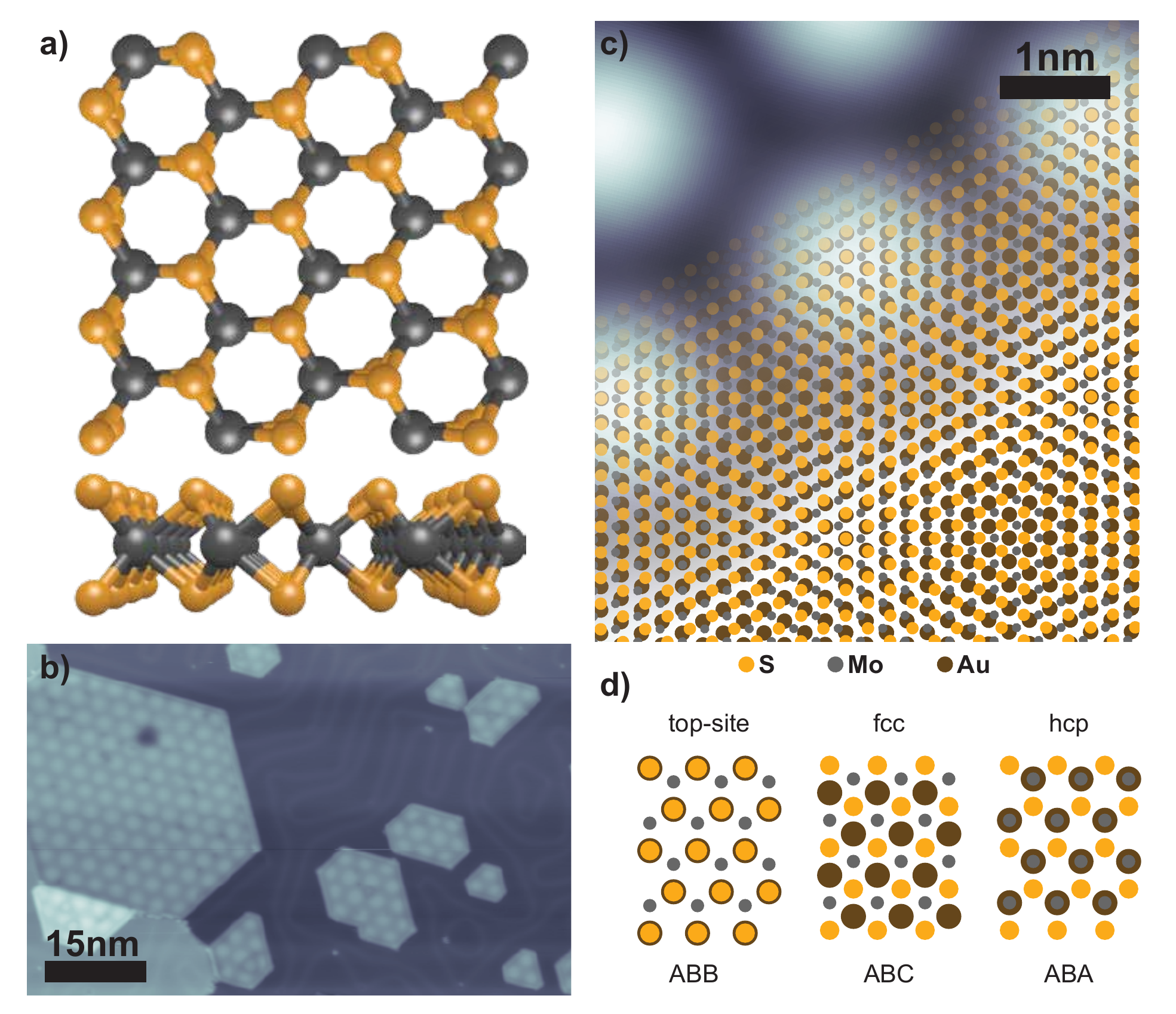}
		\caption{\textbf{(a)} Stick-and-ball model of the single layer \mos\ 2H-structure in top and side view.
\textbf{(b)} STM topography overview of \mos\ on Au(111). The \mos\ islands exhibit a moir\'e superstructure (\SI{0.8}{\volt}/\SI{100}{\pico\ampere}).
\textbf{(c)} Close-up STM image of the moir\'e superstructure (\SI{0.8}{\volt}/\SI{50}{\pico\ampere}). Superimposed a model of \mos\ on a Au(111) surface, depicting the lattice mismatch and the resulting moir\'e pattern.
\textbf{(d)} Model of the stacking order of the moir\'e superstructure at top site and the two hollow sites.}
		\label{fig:01}
	\end{center}
\end{figure*}

The above described growth procedure leads to single-layer \mos\ islands on Au(111) (Fig.~\ref{fig:01}b).
The herringbone reconstruction of the Au(111) surface is lifted by the \mos\ islands, leading to the irregular orientation of the soliton lines around the \mos\ islands~\cite{Gronborg}.
The islands appear with a characteristic hexagonal moir\'e pattern with a periodicity of about 3.3\,nm ~\cite{Bruix,Sorensen}.
This structure occurs due to the lattice mismatch of the Au substrate and the \mos\ layer, as depicted in Fig.~\ref{fig:01}c.
As a result the moir\'e unit cell contains three domains with different stacking of \mos\ and the top Au layer: one top-site and two hollow-site alignments (Fig.~\ref{fig:01}c,d).
Omitting the upper S layer, we label the lattice site of the Mo layer with "A" and the lattice site of the bottom S layer with "B", whereas the lattice site of the upper Au layer depends on the stacking domain.
When the bottom layer S atoms are on top of Au atoms, the layer stacking can be referred to as ABB, since S and Au have the same lattice site.
In the fcc-like hollow-site stacking of the bottom S layer, the Mo atom is also in a hollow-site, thus yielding an ABC stacking.
In the hcp-like hollow-site stacking of S, the Mo atoms are above the top-layer Au atoms, leading to an overall ABA stack.

\section[Characterization of stacking sequence of MoS2 layers on Au(111)]{Characterization of stacking sequence of \mos\ layers on Au(111)}
\label{stacking}

\begin{figure*}[ht]
	\begin{center}
		\includegraphics[width=0.95\textwidth]{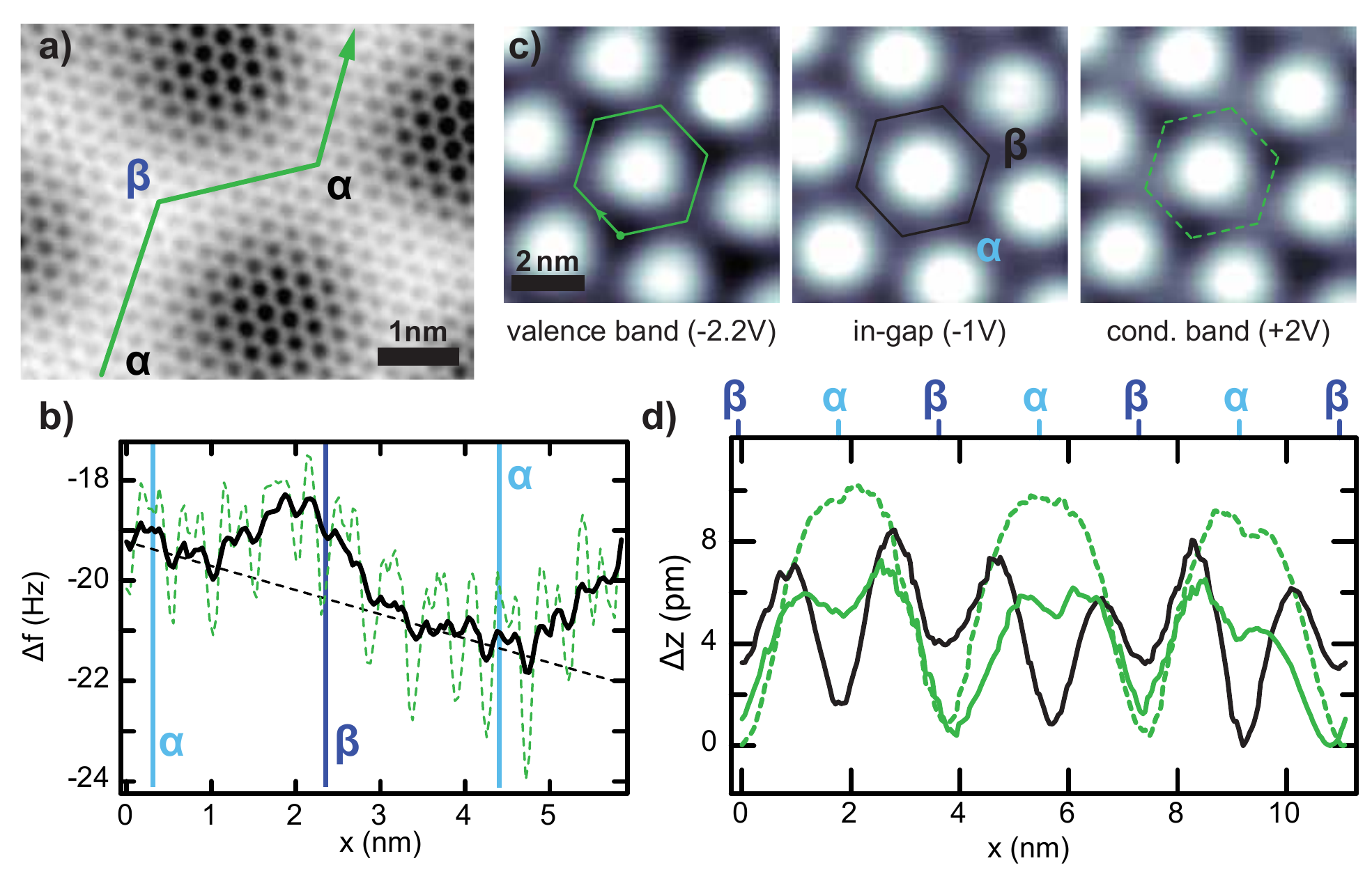}
		\caption{\textbf{(a)} Constant-height AFM frequency-shift image of \mos\ measured in the attractive regime. The dark depressions represent the upper sulfur atoms and the larger dark areas correspond to the top sites of the moir\'e pattern. The $\Delta f$ signal  along the green path is displayed in (b). Feedback opened at \SI{50}{\milli\volt}/\SI{150}{\pico\ampere} on top site and approached by \SI{210}{\pico\meter}.
\textbf{(b)} Frequency shift (green dotted) along path marked in (a) and smoothed data set (black). The black dashed line indicates an approximate linear offset in the plane. The higher absolute frequency shift, indicates hollow-site $\alpha$ to be topographically higher than hollow-site $\beta$.
\textbf{(c)} STM images of the moir\'e pattern, measured at \SI{-2.2}{\volt}/\SI{500}{\pico\ampere}, \SI{-1}{\volt}/\SI{100}{\pico\ampere} and \SI{2}{\volt}/\SI{500}{\pico\ampere}.
\textbf{(d)} Apparent height of the \mos\ along a closed path marked in (c) measured at energies in the band gap (black) and out of the gap (green). The corresponding hollow-sites are marked at the top axis.}
		\label{fig:02}
	\end{center}
\end{figure*}

The moir\'e structure found in STM images is an expression of the periodically modulated topographic height convoluted with variations in the electronic structure.
To disentangle topographic and electronic effects, we performed atomic force microscopy (AFM) measurements.
A constant-height frequency-shift image is shown in Fig.~\ref{fig:02}a, measured in the attractive force regime.
In this measurement mode, the top-layer sulfur atoms appear as depressions in the frequency shift.
The two different hollow sites are labelled $\alpha$ and $\beta$ in the following, since we cannot assign a stacking order yet.
The $\Delta f$ signal along the two hollow sites is marked in the image and displayed in Fig.~\ref{fig:02}b as dashed green line.
For better visibility a smoothed contour was added in black.
The linear offset in the graph is due to a slight angle in the constant-height plane.
Taking this linear offset into account, hollow-site $\beta$ provides about \SI{2}{\hertz} less frequency shift than compared to hollow-site $\alpha$ and is thereby topographically lower.

In the STM the situation appears reversed, when measuring at energies in the band gap of \mos, which is on Au(111) between \SI{-1.4}{\electronvolt}~\cite{Miwa} and \SI{0.5}{\electronvolt}~\cite{Cabo}.
Figure \ref{fig:02}c shows the topography of \mos\ when measured with STM at energies of the valence band (left), conduction band (right) and in the band gap (middle).
The apparent topographic heights measured along a closed path marked in figure \ref{fig:02}c are displayed in figure \ref{fig:02}d.
When measured within the band gap, hollow-site $\beta$ appears higher than $\alpha$, even though it is topographically lower, as shown by AFM.
At higher energies outside of the band gap, hollow-site $\alpha$ appears higher again.

Comparing the STM and AFM results yields information about the hybridization of \mos\ with the Au surface.
Small deviations in the distance between \mos\ and substrate, lead to a stronger or weaker hybridization and, thus, changes the in-gap STM signal.
Even though hollow-site $\beta$ is closer to the surface, the stronger hybridization provides a larger tunneling current, causing it to appear higher in STM.
In the extreme case of a missing Au layer under a small area of \mos\ -- due to a vacancy island of a few nanometers -- that area already appears completely transparent for the STM~\cite{Pits}.

\begin{figure*}[ht]
	\begin{center}
		\includegraphics[width=0.8\textwidth]{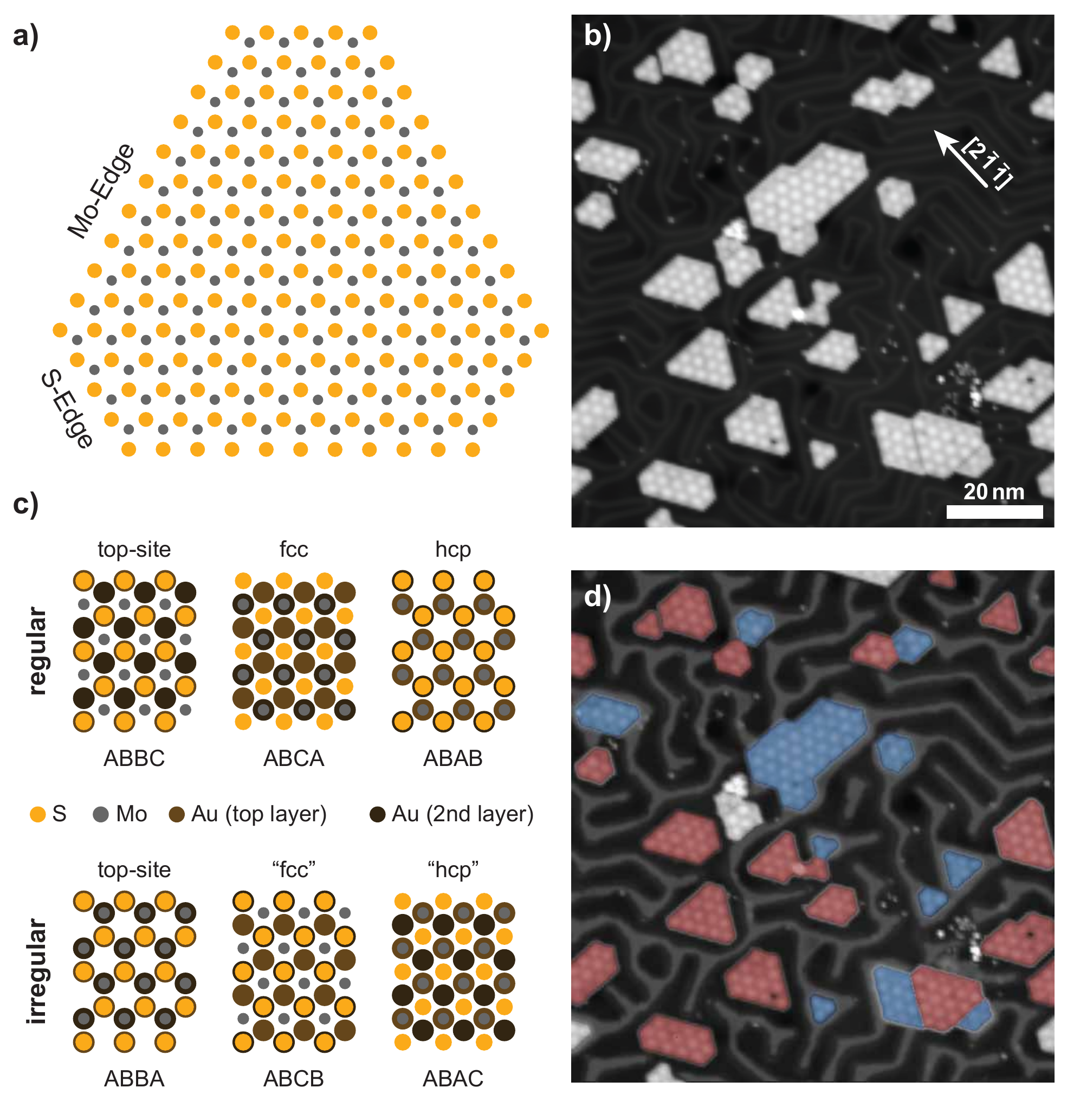}
		\caption{\textbf{(a)} Schematic model of a small \mos\ islands providing different types of edges.
\textbf{(b)} STM overview of \mos\ islands on Au(111) (\SI{0.8}{\volt}/\SI{10}{\pico\ampere}).
\textbf{(c)} Model of the stacking order of the moir\'e superstructure at top-site and the two hollow-sites including the second layer of Au(111).
In the "regular" orientation the two hollow sites stack in a normal fcc or hcp order at the Mo-S-Au-Au interface.
\textbf{(d)} Same as (b) with the two different orientations of the islands marked in red and blue, respectively. The orientation of the unmarked islands could not be determined. The hcp domains of Au(111) have been colored in light grey.}
		\label{fig:03}
	\end{center}
\end{figure*}

In order to assign the two different hollow sites to a certain stacking sequence, the orientation of the \mos\ islands have to be known.
Small islands of \mos\ on Au(111) often appear as triangles or truncated triangles, where the longer edges are assumed to be the so-called Mo-edges \cite{Bollinger,Nielsen} as depicted in Fig.~\ref{fig:03}a.
Following this assumption, the lattice orientation of the \mos\ islands can be determined and compared to the orientation of the different hollow-sites.
Thereby, the hollow-site $\alpha$ would correspond to the ABC stacking and $\beta$ to the ABA stacking sequence as is sketched in Fig.~\ref{fig:01}d.

For a complete characterization of the adsorption structure on the Au(111) surface, also the second layer of the Au(111) crystal has to be taken into account. With respect to the first two layers of Au, there are two orientations in which a \mos\ island can grow. These are sketched in Fig.~\ref{fig:03}c and labeled as regular and irregular stacking.
In the "regular" orientation all four layers of the Mo-S-Au-Au interface are stacked in a normal fcc or hcp fashion at the two hollow sites, which can be associated to an ABCA and ABAB stacking, respectively.
In contrast, the "irregular" orientation leads to a mixed stacking with ABCB and ABAC sequence, which is not fully compliant with the fcc or hcp nomenclature.

If there is a preferred stacking of \mos\ on the Au(111) crystal, it will be observable in the ratio of the two different orientations.
An overview of \mos\ islands on Au(111) is displayed in Fig.~\ref{fig:03}b and in Fig~\ref{fig:03}d, where their orientations are marked by red or blue color.
The orientation of larger islands without triangular shape was determined by the orientations of their hollow sites (see Fig.~\ref{fig:02}c).
Since the first two layers of the Au(111) surface are of interest, the surface reconstruction has to be taken into account here.
This herringbone structure is a $22 \times \sqrt{3}$ reconstruction, where the upper layer of the Au(111) surface alternates between fcc and hcp stacking~\cite{Hove, Barth}.
By fitting a hexagonal grid to an image with atomic resolution of the herringbone reconstruction, the orientation of the Au crystal was determined and is depicted in Fig.~\ref{fig:03}b.
The hcp domains of the Au(111) surface are marked in light grey in Fig.~\ref{fig:03}d.

The \mos\ islands in Fig.~\ref{fig:03}d with orientation marked red, are all lying on the fcc domain of the herringbone reconstruction.
In contrast, the blue marked islands, which provide a reversed orientation, are all located in the hcp domain.
This behaviour can also be seen in the STM images of other published data~\cite{Nielsen,Lauritsen}.
Thus, both orientations provide the same stacking of the Mo-S-Au-Au interface, indicating there is an energetically favourable stacking of \mos\ on Au(111).
Assuming again the long edges to be Mo-edges, the preferred orientation is the regular stacking as depicted in Fig.~\ref{fig:03}c.

Summarizing this part, we have shown that the \mos\ islands all exhibit the very same orientation and overall the same atomic interface structure with respect to the Au(111) surface.
We were also able to assign the two hollow sites $\alpha$ and $\beta$ to the fcc and hcp stacking order, respectively.
With the moir\'e structure providing differences in the local atomic structure, we now turn to the question, whether the electronic structure is affected by the different atomic configurations in the top sites and hollow sites.

\section[Electronic structure of MoS2 on Au(111)]{Electronic structure of \mos\ on Au(111)}

\begin{figure*}[ht]
	\begin{center}
		\includegraphics[width=0.8\textwidth]{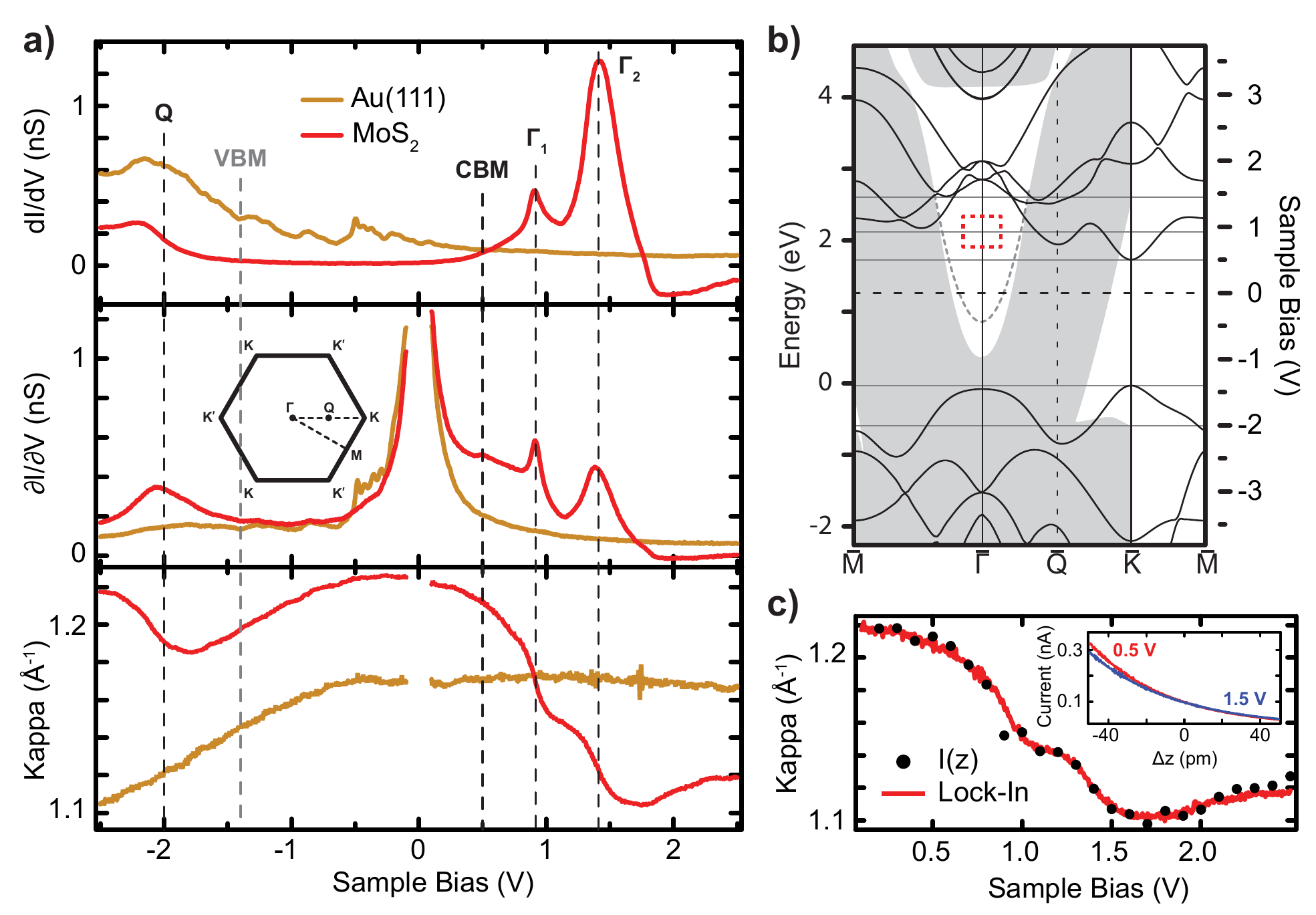}
		\caption{\textbf{(a)} Upper panel: \iv\ spectra of \mos\ (red) and Au(111) (gold). Feedback opened at \SI{2.5}{\volt}/\SI{500}{\pico\ampere} on \mos\ and \SI{2.5}{\volt}/\SI{200}{\pico\ampere} on Au(111).
Middle panel: \zv\ spectra taken with feedback on at \SI{100}{pA}.
Bottom panel: \iz\ signal taken simultaneously with \zv\ spectra. The signal was scaled to equal the decay constant $\kappa$ as shown in (c).
Dotted lines mark the position of VBM and CBM, as well as the three prominent resonances at \SI{-2}{\volt}, \SI{0.9}{\volt} and \SI{1.4}{\volt}.
\textbf{(b)} Calculated band structure of free-standing single-layer \mos.
On the energy scale on the left $E_F$ is rigidly shifted to the VBM of the calculation.
The energy scale on the right is shifted and rescaled to the energies of VBM and CBM of \mos /Au(111) according to ARPES measurements~\cite{Miwa,Cabo}.
Along \mpoint \gp\ and \gp \kp\ the band structure (grey) and surface state (dotted grey line) of Au(111) is sketched in background~\cite{Takeuchi}. The red box indicates the expected location of the $\Gamma_1$ state in the Brillouin zone.
\textbf{(c)} Decay constant measured by lock-in (red line) and by $I(z)$ measurements (black dots). Inset: Two $I(z)$ spectra measured at \SI{0.5}{\volt} (red) and \SI{1.5}{\volt} (blue).}
		\label{fig:04}
	\end{center}
\end{figure*}

In order to resolve the local electronic structure of \mos\ we first employ the most common spectroscopic mode of STM.
This method consists of recording constant-height \iv\ spectra as depicted in the uppermost graph of Fig.~\ref{fig:04}a.
The spectrum was taken at the top-site configuration of the moir\'e pattern and shows three prominent features.
There is a shoulder at a bias voltage of \SI{-2}{\volt}, and there are two peaks at \SI{0.9}{\volt} and \SI{1.4}{\volt}.
Neither of the three features matches the expected energies from ARPES measurements of the valence band maximum (VBM) at \SI{-1.4}{\electronvolt}~\cite{Miwa} or conduction band minimum (CBM) at \SI{0.5}{\electronvolt}~\cite{Cabo}, marked by vertical dashed lines.
Since they are located at the \kp\ point of the BZ (see Fig.~\ref{fig:04}b), they are not easily detectable in a \iv\ spectrum.
This can be explained by STS being mainly sensitive to states with small in-plane momentum $k_\parallel$ around the \gp\ point.
The tunneling current is strongly distance dependent, with an exponential decay $I(z) \propto \exp{\left ( -2\kappa z \right )}$ ~\cite{Tersoff1983, Tersoff1985}.
The decay constant $\kappa=\sqrt{2m\Phi/\hbar^2 + k_\parallel^2}$ depends on the energy barrier for tunneling $\Phi$, as well as on $k_\parallel$ of the probed state.
Thus, states with a large $k_\parallel$ component -- e.g. the VBM and CBM at the edge of the BZ -- have a large decay constant and therefore do not contribute significantly to the total tunneling current at large tip-sample distances.
Hence, it is not surprising to not find signatures of the VBM and CBM in the \iv\ spectrum.

In constant-current \zv\ spectra the tunneling barrier height is adjusted by a feedback loop, which regulates the current to a constant value.
A spectrum recorded in this mode is shown in Fig.~\ref{fig:04}a in the middle panel.
The tip-height adjustment enhances the sensitivity to states with larger $k_\parallel$ component at energies smaller than the band gap at \gp.
With this method, we detect a small feature at \SI{0.5}{\electronvolt}, which agrees with the expected value for the CBM from ARPES measurements~\cite{Cabo}.

Recently, Zhang \textit{et al.} reported an STS-based method using the enhanced sensitivity of \zv\ spectra, in order to measure the $k_\parallel$ of states and, thus, determine their position in the BZ.
By modulating the tip height $z$, they were able to measure the \iz\ signal with a lock-in amplifier, as long as the modulation is faster than the constant-current feedback.
The decay constant is then determined by $\kappa = -(\mathrm{d}I/\mathrm{d}z)/2I_0$, where $I_0$ is the set-point current.

We adapted this method by measuring the \zv\ and \iz\ signal simultaneously.
For the \iz\ measurement the tip was oscillated with \SI{834}{\hertz} at an amplitude of \SI{20}{\pico\meter}.
Control measurements showed no significant cross\-talk between the \zv\ and \iz\ signals.
In order to calibrate the lock-in \iz\ signal, we also recorded $I(z)$ spectra at different bias voltages.
These I(z) curves were fitted by $I(z)=\exp{\left ( -2\kappa z \right )}$ (see inset in Fig.~\ref{fig:04}c).
The extracted $\kappa$ values were used to rescale the lock-in signal of the \iz\ measurement.

The rescaled \iz\ spectrum for \mos\ on Au(111) is displayed in the bottom panel of Fig.~\ref{fig:04}a.
The reference spectrum, taken on the Au(111) surface, shows a nearly constant $\kappa$ value of \SI{1.17}{\per\angstrom} at positive bias voltages.
As sketched in the background of Fig.~\ref{fig:04}b, Au provides a projected bulk band gap at \gp\ around and above $E_F$.
This hosts a Shockley surface state, which dominates in the tunneling current.
Due to the parabolic dispersion of the surface state, the corresponding $k_\parallel$ value increases with energy, shifting $\kappa$ to higher values.
In contrast, a larger bias voltage lowers the energy barrier for tunneling $\Phi=\Phi_0 -e \lvert V \rvert /2$, with $\Phi_0$ being the energy barrier without applied bias.
These two effects cancel each other and lead to a nearly constant $\kappa$.

At negative bias voltages below \SI{-1}{\volt}, Au exhibits occupied bands around \gp.
Consequently, the main contribution to the tunneling current originates from tunneling out of these states, leading to a decrease of $\kappa$.
Please note, the measured $\kappa$ value is the averaged decay constant of the whole tunnel current, not of single states.
Hence, the $\kappa$ signal increases or decreases with the bias voltage when tunneling into a state, with the steepest slope of $\kappa$ at the maximum in the \zv\ spectrum.
For a good estimation of $k_\parallel$ the bias voltage should be larger than the corresponding state, in order to have the highest contribution of tunneling current into that state.

At small bias voltages the decay constant measured on \mos\ is \SI{1.225}{\per\angstrom}, which is considerably larger than on Au.
This indicates that the \mos\ layer does not exhibit states at \gp\ in this energy range and a higher energy barrier of the tunneling process than Au(111).
The latter agrees with the observation of an $n$-doped \mos\ layer on Au(111)~\cite{Bruix}.

With increasing bias voltage, $\kappa$ decreases on \mos.
A pronounced drop of $\kappa$ occurs at \SI{0.9}{\volt} and \SI{1.4}{\volt}, indicating that tunneling at these bias voltages is accompanied by a drastic change in the $k_\parallel$ component.
This evidences the onset of tunneling into states at different parts of the BZ and suggests that the tunneling spectra may be used for identifying characteristic points in the BZ.
In particular, the drop of $\kappa$ of \mos\ below the value measured on Au(111) at \SI{1}{\volt} shows that \mos\ states are located in the projected bulk band gap of Au.

To identify the relevant bands in the tunneling spectra, we now compare the location of the sudden change in $\kappa$ with band-structure calculations of free-standing \mos\ (Fig.~\ref{fig:04}b).
They reflect the direct band gap at \kp\ and an increased band gap at \gp.
The onset of unoccupied bands lies at approximately \SI{1.8}{\electronvolt} at \gp.
We suggest to associate these band onsets with the peak at \SI{1.4}{\volt} in the \iv\ spectra.
This assignment is corroborated by the drop in $\kappa$, which is in agreement with tunneling into bands at \gp.
In the following, this state will be labeled as $\Gamma_2$.
At bias voltages larger than \SI{2}{\volt} the decay constant increases again, in agreement with the band structure calculations, where the parallel momentum of the bands increases again.

Assuming the current at \SI{1.8}{\volt} is dominated by tunneling into $\Gamma_2$, i.e., $k_{\parallel,2}=\SI{0}{\per\angstrom}$, a rough estimation would be that $\kappa_2=\sqrt{2m\Phi}/\hbar$.
By comparing $\kappa_2$ with the $\kappa_i$ of a different state, the parallel momentum of this state can be estimated using
$k_{\parallel,i}^2 \approx \kappa_i^2 - \kappa_2^2 + \frac{me}{\hbar^2} \left ( \lvert V_i \rvert - \lvert V_2 \rvert \right )$.

Depending on whether the total decay constant is increasing or decreasing by tunneling into the state, this estimation yields a lower or a upper boundary value, respectively.
Applied to the state at \SI{0.9}{\volt}, it results in a parallel momentum of $k_\parallel \leq \SI{0.08}{\per\angstrom}$, whereas the \kp\ point is located at \SI{1.325}{\per\angstrom}~\cite{Boker}.
According to these results, there has to be a state located at the \gp\ point approximately \SI{0.4}{\electronvolt} above the CBM, marked by a red box in figure \ref{fig:04}b.
This state will be labeled as $\Gamma_1$ in the following and may originate from hybridization of Au and \mos\ states or from the formation of an interface state.
Please note, that other calculations in literature, which include the underlying Au substrate, do not predict a \mos\ state \SI{0.4}{\electronvolt} above the CBM either~\cite{Bruix}.

At negative bias voltage the decay constant of \mos\ first decreases in a similar fashion as the one of Au, indicating that the current in this regime is dominated by tunneling into the Au states.
We observe a state in the \zv\ spectra at \SI{-2}{\volt}, which is accompanied by an increase of $\kappa$ up to \SI{1.22}{\per\angstrom}.
Since the largest part of the tunneling at \SI{-2.5}{\volt} happens from this state, we can estimate its parallel momentum with $k_\parallel \approx \SI{0.6}{\per\angstrom}$.
This corresponds to a location at \qp\ in the BZ. Accordingly, the state will be referred to as $Q$ in the following.
For free-standing \mos\ a band at \gp\ would have been expected at energies close to the VBM and crossing \qp\ at energies further away from the Fermi level (see figure \ref{fig:04}b).
However our measurements do not indicate any state at \gp.
This observation is in agreement with ARPES results~\cite{Miwa} and has been explained by hybridization of the \mos\ states with the $d$-band continuum states of Au~\cite{Bruix}.
We note that we cannot identify the VBM, which we assign to the presence of the hybridized states at the same energy around \gp, which dominate the tunneling current.

\section{Influence of the moir\'e pattern on the local spectra}

Our previous results evidence the different contributions of states at different points in the BZ to the tunneling current.
Admittedly, this is not the simplest and most obvious way for determining the electronic band structure of a surface.
However, STM provides unprecedented spatial resolution of the electronic properties.
Hence, we investigate now the influence of atomic-scale structural changes on the local electronic properties.

\begin{figure}[ht]
	\begin{center}
		\includegraphics[width=0.45\textwidth]{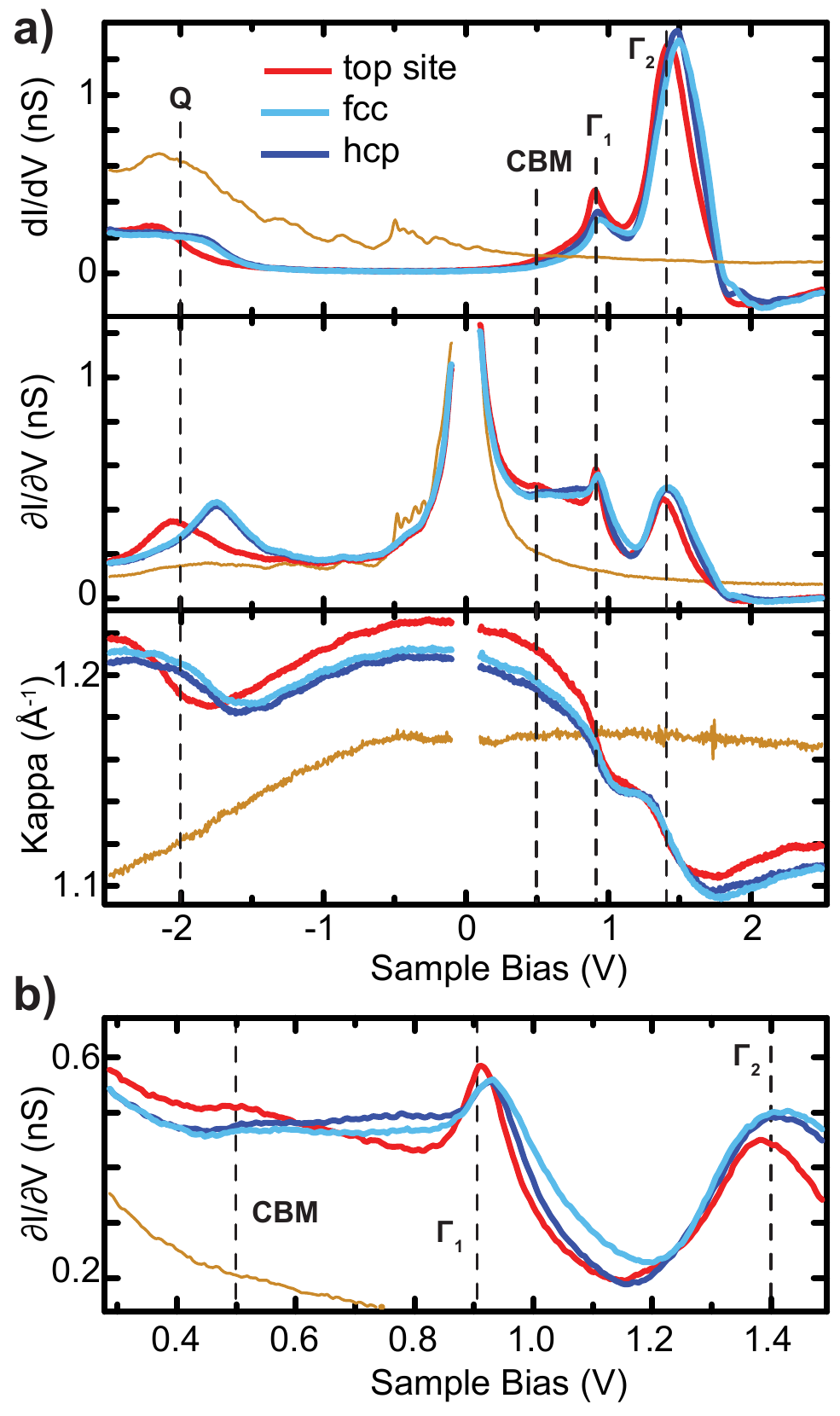}
		\caption{\textbf{(a)} Upper panel: \iv\ spectra of Au(111) (gold) and different areas of \mos\ at top site (red) and both hollow sites (light and dark blue). Feedback opened at \SI{2.5}{\volt}/\SI{500}{\pico\ampere} on \mos\ and \SI{2.5}{\volt}/\SI{200}{\pico\ampere} on Au(111).
Middle panel: \zv\ spectra taken with feedback on at \SI{100}{pA}.
Bottom panel: \iz\ signal taken simultaneously with \zv\ spectra.
\textbf{(b)} Zoom into the \zv\ spectra in the energy range of state $\Gamma_1$}
		\label{fig:05}
	\end{center}
\end{figure}

In Fig.~\ref{fig:05}a the tunneling spectra for the three different regions of the moir\'e superstructure are displayed.
The two different hollow sites are labeled fcc and hcp, according to the findings above.
The most pronounced difference in the spectra is that the state $Q$ is shifted by \SI{300}{\milli\electronvolt} towards $E_F$ for both hollow sites compared to the spectrum taken at top site.
This shift could be caused by a stronger hybridization of the \mos\ with the Au substrate or due to a more effective electronic screening~\cite{Roesner}.
The unoccupied state $\Gamma_2$ is slightly shifted away from $E_F$ by \SI{60}{\milli\electronvolt}.

In the \zv\ spectrum a distinction between all three sites can be observed around $\Gamma_1$ (Zoom shown in Fig.~\ref{fig:05}b).
In the energy range between the CBM and $\Gamma_1$ the signal of the top site differs from the two hollow-sites, whereas between $\Gamma_1$ and $\Gamma_2$ the fcc hollow site differs from the other two spectra.
Except for a low-intensity signal at \SI{2}{\volt}, all other features are nearly identical for the two hollow sites.
Thus, surprisingly, the state $\Gamma_1$ seems to be affected by the stacking order in the hollow sites.
This observation supports the assumption that the $\Gamma_1$ state at \SI{0.9}{\electronvolt} occurs due to interactions of \mos\ with the Au(111) substrate.
It is also in agreement with the absence of such a state when the \mos\ is placed on other materials with weaker interaction like graphene~\cite{Zhang}.

Calculations of \mos\ on Au(111) from literature~\cite{Bruix} predict a rigid shift of \SI{300}{\milli\electronvolt} between top site and hollow sites, whereas we observe an unequal shift of $Q$ and $\Gamma_2$.
Aside from that difference, the details of the spectra agree, \textit{i.e.} the shift of the low-intensity signal at \SI{2}{\volt} and the variation of $\Gamma_1$ between the two hollow sites agree with the predictions from Ref.~\cite{Bruix}.
The latter can also be seen in STS taken at room temperature~\cite{Sorensen}.

The tunneling decay constant $\kappa$ is not significantly affected by the  differences between the hollow sites.
The $\kappa$ spectra only reveal a difference between the hollow and top site with a reduced barrier by \SI{0.015}{\per\angstrom} within the band gap at the hollow sites.
This effect could be due to the reduced distance between \mos\ and Au(111), causing a smaller dipole moment at the interface.

\section{Conclusions}

The electronic properties of \mos\ on Au(111) deviate significantly from those of their free-standing analogues.
ARPES combined with theoretical calculations had already provided an in-depth analysis of the hybridization effects at the valence band~\cite{Bruix, Miwa}.
Our work adds information on the local scale to these earlier studies.
Using STM and AFM experiments, we are able to identify the precise stacking structure of the moir\'e pattern on the two topmost layers of Au(111).
This result does not only provide a model of the atomic structure at the \mos/Au(111) interface, it also evidences a unique growth mode.
We show that all \mos\ islands have a certain orientation on the substrate, depending on the stacking of the two top Au layers.
Assuming the so called Mo-edges to be the preferred termination of the islands, the exact stacking of \mos\ on Au(111) could be determined. Recently, Bana \textit{et al.} achieved the growth of a large, single-orientation \mos\ monolayer on Au(111)~\cite{Bana}. We speculate that the growth of such a single-orientation layer is driven by the unique stacking sequence, which we also find in the smaller islands.

Using a recently proposed method of tunneling spectroscopy and extracting the voltage-dependent tunneling decay constant of the hybrid \mos/Au(111) structure, we show that there is a state of \mos\ located at the \gp\ point around \SI{0.9}{\electronvolt}.
This state is absent in free-standing \mos\ and can be ascribed to hybridization with the underlying substrate.
We also show that this state is sensitive to the local atomic stacking configuration of the \mos\ layer on Au(111).

Our results reveal that atomic-scale changes at a \mos\ interface may significantly influence the electronic structure.
This puts stringent requirements of clean interface or a precise knowledge of defects on the design of \mos -based devices.

\section{Acknowledgments}
We gratefully acknowledge funding by the Deutsche Forschungsgemeinschaft through Sfb 658 and TRR 227.


\begin{thebibliography}{99}

\bibitem{Mak}
K.~F. Mak, C.~Lee, J.~Hone, J.~Shan, T.~F. Heinz, Atomically thin {MoS$_2$}: A
  new direct-gap semiconductor, Phys. Rev. Lett. 105~(13) (2010) 136805.
\newblock \href {http://dx.doi.org/10.1103/PhysRevLett.105.136805}
  {\path{doi:10.1103/PhysRevLett.105.136805}}.

\bibitem{Splendiani}
A.~Splendiani, L.~Sun, Y.~Zhang, T.~Li, J.~Kim, C.~Chim, G.~Galli, F.~Wang,
  Emerging photoluminescence in monolayer {MoS$_2$}, Nano Lett. 10~(4) (2010)
  1271.
\newblock \href {http://dx.doi.org/10.1021/nl903868w}
  {\path{doi:10.1021/nl903868w}}.

\bibitem{Tawinan}
T.~Cheiwchanchamnangij, W.~R.~L. Lambrecht, Quasiparticle band structure
  calculation of monolayer, bilayer, and bulk {MoS$_2$}, Phys. Rev. B 85~(20)
  (2012) 205302.
\newblock \href {http://dx.doi.org/doi.org/10.1103/PhysRevB.85.205302}
  {\path{doi:doi.org/10.1103/PhysRevB.85.205302}}.

\bibitem{Ashwin}
A.~Ramasubramaniam, Large excitonic effects in monolayers of molybdenum and
  tungsten dichalcogenides, Phys. Rev. B 86~(11) (2012) 115409.
\newblock \href {http://dx.doi.org/10.1103/PhysRevB.86.115409}
  {\path{doi:10.1103/PhysRevB.86.115409}}.

\bibitem{Komsa}
H.~Komsa, A.~V. Krasheninnikov, Effects of confinement and environment on the
  electronic structure and exciton binding energy of {MoS$_2$} from first
  principles, Phys. Rev. B 86~(24) (2012) 241201.
\newblock \href {http://dx.doi.org/10.1103/PhysRevB.86.241201}
  {\path{doi:10.1103/PhysRevB.86.241201}}.

\bibitem{Qiu}
D.~Y. Qiu, F.~H. da~Jornada, S.~G. Louie, Optical spectrum of {MoS$_2$}:
  Many-body effects and diversity of exciton states, Phys. Rev. Lett. 111~(21)
  (2015) 216805.
\newblock \href {http://dx.doi.org/10.1103/PhysRevLett.111.216805}
  {\path{doi:10.1103/PhysRevLett.111.216805}}.

\bibitem{Ugeda}
M.~Ugeda, A.~J. Bradley, S.~Shi, F.~H. da~Jornada, Y.~Zhang, D.~Qiu, W.~Ruan,
  S.~Mo, Z.~Hussain, Z.~Shen, F.~Wang, S.~Louie, M.~F. Crommie, Giant bandgap
  renormalization and excitonic effects in a monolayer transition metal
  dichalcogenide semiconductor, Nature Materials 13 (2014) 1091.
\newblock \href {http://dx.doi.org/1038/nmat4061} {\path{doi:1038/nmat4061}}.

\bibitem{Gronborg}
S.~S. Gr{\o}nborg, S.~Ulstrup, M.~Bianchi, M.~Dendzik, C.~E. Sanders, J.~V.
  Lauritsen, P.~Hofmann, J.~A. Miwa, Synthesis of epitaxial single-layer
  {MoS$_2$} on {Au(111)}, Langmuir 31~(35) (2015) 9700.
\newblock \href {http://dx.doi.org/10.1021/acs.langmuir.5b02533}
  {\path{doi:10.1021/acs.langmuir.5b02533}}.

\bibitem{Sorensen}
S.~G. S{\o}rensen, H.~G. F{\"u}chtbauer, A.~K. Tuxen, A.~S. Walton, J.~V.
  Lauritsen, Structure and electronic properties of in situ synthesized
  single-layer {MoS$_2$} on a gold surface, ACS Nano 8~(7) (2014) 6788.
\newblock \href {http://dx.doi.org/10.1021/nn502812n}
  {\path{doi:10.1021/nn502812n}}.

\bibitem{Shi}
Y.~Shi, W.~Zhoi, A.~Lu, W.~Fank, Y.~Lee, A.~L. Hsu, S.~M. Kim, K.~K. Kim, H.~Y.
  Yang, L.~Li, J.~Idrobo, J.~Kong, van der Waals epitaxy of {MoS$_2$} layers
  using graphene as growth templates, Nano Lett. 12~(6) (2012) 2784.
\newblock \href {http://dx.doi.org/10.1021/nl204562j}
  {\path{doi:10.1021/nl204562j}}.

\bibitem{Lee}
Y.~Lee, X.~Zhang, W.~Zhang, M.~Chang, C.~Lin, K.~Chang, Y.~Yu, J.~T. Wang,
  C.~Chang, L.~Li, T.~Lin, Synthesis of large-area {MoS$_2$} atomic layers with
  chemical vapor deposition, Adv. Mater. 24~(17) (2012) 2320.
\newblock \href {http://dx.doi.org/10.1002/adma.201104798}
  {\path{doi:10.1002/adma.201104798}}.

\bibitem{Zhan}
Y.~Zhan, Z.~Liu, S.~Najmaei, P.~M. Ajayan, J.~Lou, Large-area vapor-phase
  growth and characterization of {MoS$_2$} atomic layers on a {SiO$_2$}
  substrate, small 8~(7) (2012) 966.
\newblock \href {http://dx.doi.org/10.1002/smll.201102654}
  {\path{doi:10.1002/smll.201102654}}.

\bibitem{Wang}
S.~Wang, X.~Wang, J.~H. Warner, All chemical vapor deposition growth of
  {MoS$_2$}:h-bn vertical van der Waals heterostructures, ACS Nano 9~(5) (2015)
  5246.
\newblock \href {http://dx.doi.org/0.1021/acsnano.5b00655}
  {\path{doi:0.1021/acsnano.5b00655}}.

\bibitem{Roesner}
M.~R{\"o}sner, C.~Steinke, M.~Lorke, C.~Gies, F.~Jahnke, T.~O. Wehling,
  Two-dimensional heterojunctions from nonlocal manipulations of the
  interactions, Nano Lett. 16~(4) (2016) 2322.
\newblock \href {http://dx.doi.org/10.1021/acs.nanolett.5b05009}
  {\path{doi:10.1021/acs.nanolett.5b05009}}.

\bibitem{Bruix}
A.~Bruix, J.~A. Miwa, N.~Hauptmann, D.~Wegner, S.~Ulstrup, S.~S. Gr{\o}nborg,
  C.~E. Sanders, M.~Dendzik, A.~Grubi\v{s}i\'c~\v{C}abo, M.~Bianchi, J.~V.
  Lauritsen, A.~A. Khajetoorians, B.~Hammer, P.~Hofmann, Single-layer {MoS$_2$}
  on {Au(111)}: Band gap renormalization and substrate interaction, Phys. Rev.
  B 93~(16) (2016) 165422.
\newblock \href {http://dx.doi.org/10.1103/PhysRevB.93.165422}
  {\path{doi:10.1103/PhysRevB.93.165422}}.

\bibitem{Dedzik}
M.~Dedzik, A.~Bruix, M.~Michiardi, A.~S. Ngankeu, M.~Bianchi, J.~A. Miwa,
  B.~Hammer, P.~Hofmann, C.~E. Sanders, Substrate-induced
  semiconductor-to-metal transition in monolayer {WS$_2$}, Phys. Rev. B 96~(23)
  (2017) 235440.
\newblock \href {http://dx.doi.org/10.1103/PhysRevB.96.235440}
  {\path{doi:10.1103/PhysRevB.96.235440}}.

\bibitem{Miwa}
J.~Miwa, S.~Ulstrup, S.~G. S{\o}rensen, M.~Dendzik, A.~Grubi\v{s}i\'c~\v{C}abo,
  M.~Bianchi, J.~V. Lauritsen, P.~Hofmann, Electronic structure of epitaxial
  single-layer {MoS$_2$}, Phys. Rev. Lett. 114~(4) (2014) 046802.
\newblock \href {http://dx.doi.org/10.1103/PhysRevLett.114.046802}
  {\path{doi:10.1103/PhysRevLett.114.046802}}.

\bibitem{Cabo}
A.~Grubi\v{s}i\'c~\v{C}abo, J.~A. Miwa, S.~S. Gr{\o}nborg, J.~M. Riley, J.~C.
  Johannsen, C.~Cacho, O.~Alexander, R.~T. Chapman, E.~Springate, M.~Grioni,
  J.~V. Lauritsen, P.~D.~C. King, P.~Hofmann, S.~Ulstrup, Observation of
  ultrafast free carrier dynamics in single layer {MoS$_2$}, Nano Lett. 15~(9)
  (2015) 5883.
\newblock \href {http://dx.doi.org/10.1021/acs.nanolett.5b01967}
  {\path{doi:10.1021/acs.nanolett.5b01967}}.

\bibitem{Zhang}
C.~Zhang, Y.~Chen, A.~Johnson, M.~Li, L.~Li, P.~C. Mende, R.~M. Feenstra,
  C.~Shih, Probing critical point energies of transition metal dichalcogenides:
  Surprising indirect gap of single layer {WSe$_2$}, Nano Lett. 15~(10) (2015)
  6494.
\newblock \href {http://dx.doi.org/10.1021/acs.nanolett.5b01968}
  {\path{doi:10.1021/acs.nanolett.5b01968}}.

\bibitem{Giessibl}
F.~Giessibl, Atomic resolution on {Si(111)-(7x7}) by noncontact atomic force
  microscopy with a force sensor based on a quartz tuning fork, Appl. Phys.
  Lett 76~(11) (2000) 1470.
\newblock \href {http://dx.doi.org/10.1063/1.126067}
  {\path{doi:10.1063/1.126067}}.

\bibitem{QE}
P.~t. Giannozzi, Quantum espresso: a modular and open-source software project
  for quantum simulations of materials, J. Phys. Condens. Matter 21~(39) (2009)
  395502.
\newblock \href {http://dx.doi.org/10.1088/0953-8984/21/39/395502}
  {\path{doi:10.1088/0953-8984/21/39/395502}}.

\bibitem{Padilha}
J.~E. Padilha, H.~Peelaers, A.~Janotti, C.~G. Van~de Walle, Nature and
  evolution of the band-edge states in {MoS$_2$}: From monolayer to bulk, Phys.
  Rev. B 90~(20) (2014) 205420.
\newblock \href {http://dx.doi.org/10.1103/PhysRevB.90.205420}
  {\path{doi:10.1103/PhysRevB.90.205420}}.

\bibitem{Pits}
N.~Krane, C.~Lotze, J.~L{\"a}ger, G.~Reecht, K.~J. Franke, Electronic structure
  and luminescence of quasi-freestanding {MoS$_2$} nanopatches on {Au(111)},
  Nano Lett. 16~(8) (2016) 5163.
\newblock \href {http://dx.doi.org/10.1021/acs.nanolett.6b02101}
  {\path{doi:10.1021/acs.nanolett.6b02101}}.

\bibitem{Bollinger}
M.~V. Bollinger, J.~V. Lauritsen, K.~W. Jacobsen, J.~K. N{\o}rskov, S.~Helveg,
  F.~Besenbacher, One-dimensional metallic edge states in {MoS$_2$}, Phys. Rev.
  Lett. 87~(19) (2001) 196803.
\newblock \href {http://dx.doi.org/10.1103/PhysRevLett.87.196803}
  {\path{doi:10.1103/PhysRevLett.87.196803}}.

\bibitem{Nielsen}
J.~H. Nielsen, L.~Bech, K.~Nielsen, Y.~Tison, K.~P. J{\o}rgensen, J.~L. Bonde,
  S.~Horch, T.~F. Jaramillo, I.~Chrokendorff, Combined spectroscopy and
  microscopy of supported {MoS$_2$} nanoparticles, Surface Science 603~(9)
  (2009) 1182.
\newblock \href {http://dx.doi.org/10.1016/j.susc.2009.02.039}
  {\path{doi:10.1016/j.susc.2009.02.039}}.

\bibitem{Hove}
M.~A. Van~Hove, R.~J. Koestner, P.~C. Stair, J.~P. Bib\'erian, L.~L. Kesmodel,
  I.~Barto\v{s}, G.~A. Somorjai, The surface reconstructions of the (100)
  crystal faces of iridium, platinum and gold: I. experimental observations and
  possible structural models, Surface Science 103~(1) (1981) 189.
\newblock \href {http://dx.doi.org/10.1016/0039-6028(81)90107-2}
  {\path{doi:10.1016/0039-6028(81)90107-2}}.

\bibitem{Barth}
J.~V. Barth, H.~Brune, G.~Ertl, R.~J. Behm, Scanning tunneling microscopy
  observations on the reconstructed {Au(111)} surface: Atomic structure,
  long-range superstructure, rotational domains, and surface defects, Phys.
  Rev. B 42~(15) (1990) 9307.
\newblock \href {http://dx.doi.org/10.1103/PhysRevB.42.9307}
  {\path{doi:10.1103/PhysRevB.42.9307}}.

\bibitem{Lauritsen}
J.~V. Lauritsen, J.~Kibsgaard, G.~H. Olesen, P.~G. Moses, B.~Hinnemann,
  S.~Helveg, J.~K. N{\o}rskovb, B.~S. Clausenc, H.~Tops{\o}e, E.~L{\ae}sgaard,
  Location and coordination of promoter atoms in Co- and {Ni}-promoted
  {MoS$_2$}-based hydrotreating catalysts, Journal of Catalysis 249~(2) (2007)
  220.
\newblock \href {http://dx.doi.org/10.1016/j.jcat.2007.04.013}
  {\path{doi:10.1016/j.jcat.2007.04.013}}.

\bibitem{Takeuchi}
N.~Takeuchi, C.~T. Chan, K.~M. Ho, {Au(111)}: A theoretical study of the
  surface reconstruction and the surface electronic structure, Phys. Rev. B
  43~(17) (1991) 13899.
\newblock \href {http://dx.doi.org/10.1103/PhysRevB.43.13899}
  {\path{doi:10.1103/PhysRevB.43.13899}}.

\bibitem{Tersoff1983}
J.~Tersoff, D.~R. Hamann, Theory and application for the scanning tunneling
  microscope, Phys. Rev. Lett. 50~(25) (1983) 1998.
\newblock \href {http://dx.doi.org/10.1103/PhysRevLett.50.1998}
  {\path{doi:10.1103/PhysRevLett.50.1998}}.

\bibitem{Tersoff1985}
J.~Tersoff, D.~R. Hamann, Theory of the scanning tunneling microscope, Phys.
  Rev. B 31~(2) (1985) 805.
\newblock \href {http://dx.doi.org/10.1103/PhysRevB.31.805}
  {\path{doi:10.1103/PhysRevB.31.805}}.

\bibitem{Boker}
T.~B{\"o}ker, A.~Severin, A.~M{\"u}ller, C.~Janowitz, R.~Manzke, D.~Vo{\ss},
  P.~Kr{\"u}ger, A.~Mazur, J.~Pollman, Band structure of {MoS$_2$}, {MoSe$_2$},
  and {$\alpha$-MoTe$_2$}: Angle-resolved photoelectron spectroscopy and ab
  initio calculations, Phys. Rev. B 64~(23) (2001) 235305.
\newblock \href {http://dx.doi.org/10.1103/PhysRevB.64.235305}
  {\path{doi:10.1103/PhysRevB.64.235305}}.

\bibitem{Bana}
H.~Bana, E.~Travaglia, L.~Bignardi, P.~Lacovig, C.~E. Sanders, M.~Dendzik,
  M.~Michiardi, M.~Bianchi, d.~Lizzit, F.~Presel, D.~De~Angelis, N.~Apostol,
  P.~Kumar~Das, J.~Fujii, I.~Vobornik, R.~Larciprete, A.~Baraldi, P.~Hofmann,
  S.~Lizzit, Epitaxial growth of single-orientation high-quality {MoS$_2$}
  monolayers, arXiv:1802.02220.

\end{thebibliography}

\end{document}